\documentclass[pre,twocolumn,floatfix,superscriptaddress,a4paper,nofootinbib]{revtex4-2}
\usepackage{amsmath,bm,graphicx}
\usepackage{amssymb,amsfonts}
\usepackage{epstopdf}
\usepackage{mathrsfs}
\usepackage{xcolor}
\usepackage{latexsym}
\usepackage{hyperref}
\usepackage{subfigure}
\usepackage{soul}
\usepackage{bm}
\usepackage{natbib}
\usepackage{braket}
\usepackage[normalem]{ulem}

\newcommand{\pa}[1]{{\color{orange} #1}}

\begin{document}

\title{Quantum tunneling and anti-tunneling across entropic barriers}

\author{Paolo Malgaretti}
\email[Corresponding Author: ]{p.malgaretti@fz-juelich.de }
\affiliation{Helmholtz Institute Erlangen-N\"urnberg for Renewable Energy (IET-2), Forschungszentrum J\"ulich, Cauerstr. 1,
91058 Erlangen, Germany}
\author{Francesco Petiziol}
\affiliation{Technische Universit\"at Berlin, Institut für Physik und Astronomie, Hardenbergstr.~36,
10623 Berlin, Germany}
\author{Alexander Schnell}
\affiliation{Technische Universit\"at Berlin, Institut für Physik und Astronomie, Hardenbergstr.~36,
10623 Berlin, Germany}

\newcommand{\fra}[1]{\textcolor{blue}{#1}}
\newcommand{\fraq}[1]{\textcolor{blue!40!cyan}{F: #1}}

\newcommand{\canc}[1]{{\textcolor{red}{\sout{#1}}}}

\begin{abstract}
We study the dynamics of a quantum particle in a constricted two-dimensional channel and analyze how the onset of quantum corrections impacts the (semi-)classical high-temperature behaviour, as temperature is lowered. We characterize both equilibrium and non-equilibrium (transport) properties of the system, considering the case of a narrow and disorder-free channel. Counterintuitively, we find that quantum corrections do not monotonically enhance the particle current as the temperature is lowered, as naively expected from the activation of coherent tunnelling, but they rather inhibit transport at intermediate temperatures, increasing the effective free-energy barrier. We illustrate this ``anti-tunnelling'' effect numerically by computing the non-equilibrium steady-state of a quantum master equation describing the system, and confirm it analytically by adopting the Quantum Smoluchowski limit.
\end{abstract}
\maketitle

The transport of molecules, polymers, and colloids across constrictions and porous materials has attracted considerable attention due to both the theoretical challenges that it poses as well as to its relevance for technological application and bio-medical scenarios~\cite{Dagdug_book}. 
Typical examples are transport of DNA, RNA, ions, and molecules across nuclear and cellular membranes~\cite{Albers_book,Hille_book}, transport of pollutants across the soil or sieves~\cite{Zeolites_book}, and particle separation in chromatography~\cite{Chromatography_book}, among others.

From a theoretical perspective, describing transport across porous materials is a tremendous challenge, since it requires one to solve at the same time the microscopic interactions with the solid matrix as well as the mesoscopic entropic effects caused by the sequence of enlargements and constrictions along the porous material.

Theoretical effective models have been developed, where the porous material is modelled as a set of channels with varying cross-sections. 
In the case in which the channel section is varying very smoothly, it is possible to reduce the $2\mathrm{D}-3\mathrm{D}$ problem to an effective $1\mathrm{D}$ problem whose geometry is captured via the local equilibrium free energy. Such an approach, called Fick-Jackobs approximation~\cite{Zwanzig1992,Reguera2001,Malgaretti2013}, has been fruitfully exploited to study the transport of point-particles~\cite{Dagdug2017,Kalinay2020}, polymers~\cite{Malgaretti2019,Malgaretti2023}, and (active) colloids~\cite{Marconi2015,Reguera2012,Malgaretti2019}.

Transport across energy barriers whose physical origin does not stem from steric interaction with a potential barrier, but rather from `softer' Hamiltonian interactions, is also a very relevant and classic scenario in the quantum realm. For example, point contacts are used to connect macroscopic particle reservoires~\cite{Brantut12,Brantut13,husmann_connecting_2015,lebrat_quantized_2019}, light potentials are used to trap (e.g.~via optical tweezers \cite{Gieseler21}) and control cold atoms~\cite{wieman_atom_1999,Norcia18,Pagano19,Spar22,Chew24,Bluvstein24}, including Bose-Einstein condensates~\cite{bloch_non-equilibrium_2022}, and local variations in the effective friction have been investigated \cite{bridge_quantum_2024}. Geometric confinement can further enhance quantum interference effects in electron transport, as widely explored in semiconductor nanostructures~\cite{Beenakker1991}.

\begin{figure}
    \centering
    \includegraphics[scale=0.05]{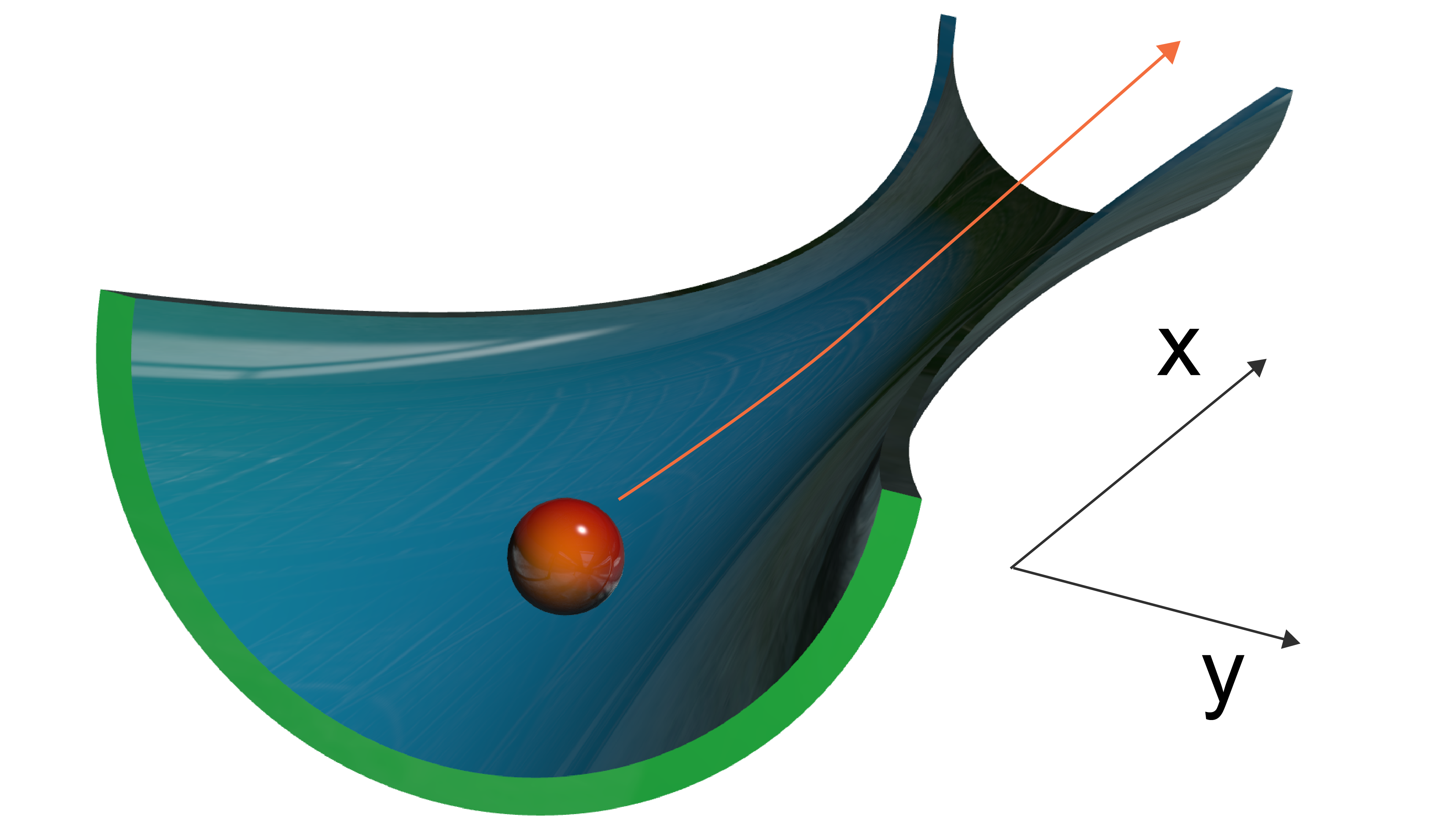}
    \caption{
    Cartoon of the potential characterized by a steep increase along the $y$ direction and a shallow variation along the $x$ direction.}
    \label{fig:scheme}
\end{figure}

Here we aim at bridging the two scenarios -- classical and quantum -- by investigating how quantum corrections affect equilibrium and transport properties around the semi-classical high-temperature limit for a particle in a two-dimensional constricted channel. In particular, we study an ideal gas particle which is freely diffusing in a potential well whose stiffness gently varies along the longitudinal direction (see Fig.~\ref{fig:scheme}) and we characterize both its equilibrium as well as non-equilibrium properties.

In order to compare to classical systems (for which the dynamics is governed by the local equilibrium free energy) we compute the local equilibrium free energy, $\mathcal{F}$ and, in particualr, the free energy difference, $\Delta \mathcal{F}\equiv \mathcal{F}(L/2)-\mathcal{F}(0)$ between the maximum and minimum of the free energy. 
Interestingly, when the potential is varying smoothly, our results show that the first quantum corrections at high temperature lead to an increase in the free energy barrier $\Delta \mathcal{F}$. This is counterintuitive, since one would expect that quantum corrections may decrease the effective barrier due to the activation of coherent tunneling, which we indeed observe eventually at lower temperatures. 
We capture this effect, which we will refer to as ``anti-tunneling", analytically by extending the Fick-Jackobs approximation to the case of a Quantum Smoluchowki equation accounting for the leading correction at high temperature. The success of this analytical approach is benchmarked against numerical solution of the quantum equation of motion of the open quantum systems.
The non-monotonic behavior of the free energy persists also in the case of transport as we observe by both numerical and analytical results.

\section{Model}

Here we study a point particle in a potential, $U(x,y)$, that is confining along the $y$ direction and periodic along the $x$ direction. The system in canonical local equilibrium with a bath at temperature $T$. We are interested in characterizing the onset of the first quantum corrections at high temperature for which standard quantum Smoluchowski equations have been derived for the case of a $1\mathrm{D}$ potential by expanding about the classical case~\cite{ankerhold_strong_2001,Hanggi2004,luczka_diffusion_2005,coffey_wigner_2007,Hanggi_RevModPhys} 
\begin{equation}
\dot{p}(y,t)=\frac{D_{cl}}{L^{2}}\partial_{y}\left[\partial_{y}\left[D_{qm}(y)p(y,t)\right]+p(y,t)\beta\partial_{y}U(y)\right],\label{eq:Smol}
\end{equation}
where 
\begin{equation}
D_{qm}(y)=\frac{1}{1-2\Lambda\partial_{y}^{2}\beta U(y)}.
\end{equation}
and $y$ is the dimensionless position of the particle. 
The small parameter of the expansion is 
\begin{align}\label{eq:def-lambda}
\Lambda = \frac{1}{48 \pi} \frac{\lambda_T^2}{L_y^2}\,,\quad\quad & \lambda_T^2=\frac{2\pi \hbar^2 \beta}{m},
\end{align} 
with $\lambda_T$ the (de Broglie) thermal wavelength,  and $\beta=(k_BT)^{-1}$ is the inverse thermal energy. $L_y$ corresponds to a typical length scale in the $y$ direction that we will specify later in the manuscript for the model systems analyzed.  In particular, $\Lambda = 0$ corresponds to the classical limit, whereas $\Lambda\rightarrow \infty$ corresponds to the zero-temperature, deep quantum regime. 
For such a case the quantum corrections affect both the effective potential as well as the diffusion coefficient in the Smoluchowski equation which acquires a dependence on the local potential. 
The equilibrium solution of Eq.~\eqref{eq:Smol}, up to order $\mathcal{O}\left(\Lambda^{2}\right)$, reads
\begin{equation}
p_{eq}(y)=N_{0}e^{-\beta U(y)}\left(1-2\Lambda\beta\partial_{y}^{2}U(y)+\Lambda\beta\left(\partial_{y}U(y)\right)^{2}\right).
\end{equation}
Here, we generalize the $1$D model of Refs.~\cite{ankerhold_strong_2001,Hanggi2004,luczka_diffusion_2005,coffey_wigner_2007} to the $2\mathrm{D}$ case by introducing the diffusion tensor 
\begin{align}
\mathbf{D}_{qm}\equiv \left[ \begin{array}{cc}
\frac{1}{1-2\Lambda L_y^2\partial_x^{2}\beta U(x,y)}
 & 0\\
0 & \frac{1}{1-2\Lambda L_y^2\partial_y^{2}\beta U(x,y)}
\end{array}\right].
\end{align}
We will check this ansatz a posteriori by comparing the predictions of our model to the numerical solution of the relevant quantum-mechanical equation of motion describing the system in contact with a thermal bath. 
According to our ansatz, the $2\mathrm{D}$ quantum Smoluchowski equation reads
\begin{equation}
\dot{p}=D_{cl}\nabla\cdot\left[\nabla\cdot\left[\mathbf{D}_{qm}(x,y)p\right]+p\beta\nabla U(x,y)\right]. \label{eq:Smol-2D}
\end{equation}
As observed in many cases for classical systems~\cite{Zwanzig1992,Reguera2001,Malgaretti2013,Dagdug2017,Kalinay2020,Malgaretti2019,Malgaretti2023,Marconi2015,Reguera2012,Malgaretti2019} insight into the interplay between the geometrical confinement and the particle dynamics can be obtained by 
assuming a length scale separation, $L_{y}\ll L_{x}$, 
between the typical length scale associated to the confining potential along the $y$- and the $x$-direction. At leading order in $\partial_{x}\beta U$, a factorization ansatz of the following form for the solution of Eq.~\eqref{eq:Smol-2D}  is justified~\cite{Zwanzig1992,Reguera2001,Malgaretti2013} 
\begin{align}
p(x,y,t)=\rho(x,t)\frac{1}{L_y}\dfrac{e^{-\beta\psi(x,y)}}{e^{-\beta A(x)}},
\label{eq:ansatz_2D}
\end{align}
where 
\begin{align}
e^{-\beta\psi}=e^{-\beta U}\left[1-2\Lambda L_y^2\partial_y^{2}\beta U+\Lambda L_y^2(\partial_{y}\beta U)^{2}\right]
\end{align}
is the equilibrium distribution~\cite{ankerhold_strong_2001,Hanggi2004,luczka_diffusion_2005,coffey_wigner_2007,Hanggi_RevModPhys}, to order $\mathcal{O}(\Lambda),$ along the transverse (i.e.~$y$-) direction and
\begin{align}
 A(x)=-\frac{1}{\beta}\ln\left[\frac{1}{L_y}\int_{-\infty}^{\infty}e^{-\beta\psi(x,y)}dy\right]
\label{eq:def-A}
\end{align}
is the local equilibrium free energy.  
Substituting the ansatz into
the quantum Smoluchowski equation leads to 
\begin{align}
\dfrac{\dot{\rho}(x,t)}{L_y}\dfrac{e^{-\beta\psi(x,y)}}{e^{-\beta A(x)}}  =  & D_{cl}\nabla\cdot\left[\nabla\cdot\left[\mathbf{D}_{qm}(x,y)\frac{\rho(x,t)}{L_y}\dfrac{e^{-\beta\psi(x,y)}}{e^{-\beta A(x)}}\right]\right.\nonumber\\
&\left. +\frac{\rho(x,t)}{L_y}\dfrac{e^{-\beta\psi(x,y)}}{e^{-\beta A(x)}}\beta\nabla U(x,y)\right].
\end{align}
Integration in the transverse direction leads to vanishing contribution of the flux along $y$. Moreover, at leading order in $\partial_x U$, we can disregard the spatial dependence of the diffusion coefficient along the $x$ direction~\cite{Zwanzig1992,Reguera2001}. Accordingly, Eq.~\eqref{eq:Smol-2D} reduces to
\begin{align}
\dot{\rho}(x,t)  = & D_{cl}\partial_{x}\left\{ \partial_{x}\int_{-\infty}^{\infty}\rho(x,t)\dfrac{e^{-\beta\psi(x,y)}}{e^{-\beta A(x)}}\frac{dy}{L_y}\right.\nonumber\\
&\left. +\int_{-\infty}^{\infty}\rho(x,t)\dfrac{e^{-\beta\psi(x,y)}}{e^{-\beta A(x)}}\beta\partial_{x}U(x,y)\frac{dy}{L_y}\right\} 
\end{align}
which, since $\rho$ does not depend on $y$ can be rewritten as 
\begin{align}
\dot{\rho}(x,t) = D_{cl}\partial_{x}\left\{ \partial_{x}\rho(x,t)+\rho(x,t)\beta \partial_{x}\mathcal{F}(x)\right\}, \label{eq:FJ_qm}
\end{align}
where 
\begin{align}
\partial_{x}\mathcal{F}(x) & =  
\int_{-\infty}^{\infty}\dfrac{e^{-\beta\psi(x,y)}}{e^{-\beta A(x)}}\partial_{x}U(x,y)\frac{dy}{L_y}
\end{align}
and $\mathcal{F}(x)$ the effective $1\mathrm{D}$ potential governing the dynamics along the longitudinal direction. 
At linear order in $\Lambda$ (see Appendix \ref{sec:app-derivFLamb}), and accounting for the fact that $\partial_{x}A(x)\partial_{x}^{2}U$, is a higher order contribution in $\beta \partial_{x} U$ we have 
\begin{align}
\beta\partial_x\mathcal{F}(x)\simeq\beta \partial_xA_0(x)+\Lambda \beta\partial_x\mathcal{F}_\Lambda(x),
\label{eq:F_final}
\end{align} 
where $A_0(x)$ is the classical contribution~\cite{Zwanzig1992,Reguera2001,Malgaretti2013} (i.e.~for $\Lambda=0$) and  
\begin{align}\label{eq:F-Lambda}
\beta & \partial_x\mathcal{F}_\Lambda(x) = \beta\partial_x A_0(x)L_y^2\int dx\int\limits_{-\infty}^{\infty}\dfrac{e^{-\beta U}}{e^{-\beta A_{0}}}\partial_{y}^{2}\beta U\frac{dy}{L_y}+\nonumber\\
&
+L_y^2\!\!\int \!dx\!\! \int\limits_{-\infty}^{\infty}\!\!\dfrac{e^{-\beta U}}{e^{-\beta A_{0}}}\left[\left(\beta \partial_y U\right)^2-\beta\partial^2_{y}U\right]\beta\partial_x U\frac{dy}{L_y}
\end{align}
is the correction to the local equilbrium free energy proportional to $\Lambda$. It is interesting to note that for the classical limit, $\Lambda =0$, we have $\mathcal{F}(x) = A_0(x)$ and hence the effective potential can  directly be read from the factorization, Eq.~\eqref{eq:ansatz_2D}, and its normalization, Eq.~\eqref{eq:def-A}. In contrast, for the quantum case, $\Lambda\neq 0$, the effective potential has a more involved functional form, Eq.~\eqref{eq:F-Lambda}, which is due to the fact that the effective force along the transverse direction, $\partial_x U$, differs from the derivative of the exponent of the Boltzmann weight, $\partial_x \psi$. 
The steady state solution of Eq.~\eqref{eq:FJ_qm} reads 
\begin{equation}
\rho(x)=e^{-\beta\mathcal{F}(x)}\left[-\frac{J}{D_{cl}}\int_{x_{1}}^{x} e^{\beta\mathcal{F}(x')}dx'+\Pi\right],
\end{equation}
where $x_{1}$ is an arbitrary position, $J$ is the current and $\Pi$ is an integration constant. Both $J$ and $\Pi$ are determined by the boundary conditions.
In the next sections we use this general result to discuss both the equilibrium as well as the transport properties of the system.

\section{Equilibrium }

At thermodynamic equilibrium, the current vanishes, $J=0$. Accordingly, the steady state solution reduces to 
\begin{equation}
\rho(x)=\Pi e^{-\beta\mathcal{F}(x)},
\label{eq:rho_eq}
\end{equation}
where $\Pi$ is determined by the normalization condition
\begin{align}
    \int_0^L \rho(x) dx = \rho_0 L e^{\beta(\mu-\mu_0)},
\end{align}
where $\mu$ is the chemical potential and $\rho_0$ is the value of the density for $\mu=\mu_0$.

In order to further inspect the dependence of the effective free energy on $\Lambda$, in the following we specialize to the case of a modulated harmonic potential of the form 
\begin{subequations}\label{eq:pot}
\begin{align}
 U(x,y) & =\frac{1}{2} k(x)y^{2}, \\
k(x) & =k_{0}\left[1+k_{1}\cos\left(\frac{2\pi x}{L}\right)\right],
\end{align}
\end{subequations}
where $L$ is the period of the potential. In the case of harmonic potential, it is natural to introduce a (finite-temperature) transverse length scale $L_y = \sqrt{2k_BT/k_0}$, which captures the effective range of values explored due to thermal fluctuations. 
For a harmonic potential, up to first order in $\Lambda$, $\rho(x)$ reads
\begin{align}
    \rho(x) = \rho_0 e^{\beta(\mu-\mu_0)}( \rho_{cl}(x)+\Lambda\rho_\Lambda(x))
    \label{eq:corr-rho-lambda}
\end{align}
with 
\begin{align}
    \rho_{cl}(x) &= e^{-\beta A_0(x)}=\sqrt{\pi\frac{k_0}{ k(x)}},\\
    \rho_\Lambda(x) & = -\!\int_0^L \!\dfrac{e^{-\beta U(x,y)}}{e^{-\beta A_0(x)}} L_y^2 \partial_y^{2}\beta U(x,y) dy =-2\frac{k(x)}{k_0}.
    \label{eq:rho-lambda}
\end{align}  
Similarly, after some algebra (see Appendix~\ref{app:DF-harm}), for the free energy we get
\begin{align}\label{eq:free_en-harm}
\beta \mathcal{F}^{hrm}(x)&= \frac{1}{2}\ln\frac{k(x)}{k_0}+2\Lambda \frac{k(x)}{k_0}\,.
\end{align}
Equation \eqref{eq:free_en-harm} shows that the free energy has two contributions: the classical (entropic) contribution plus a linear correction in $\Lambda$. 

\begin{figure}
\includegraphics[scale=0.50]{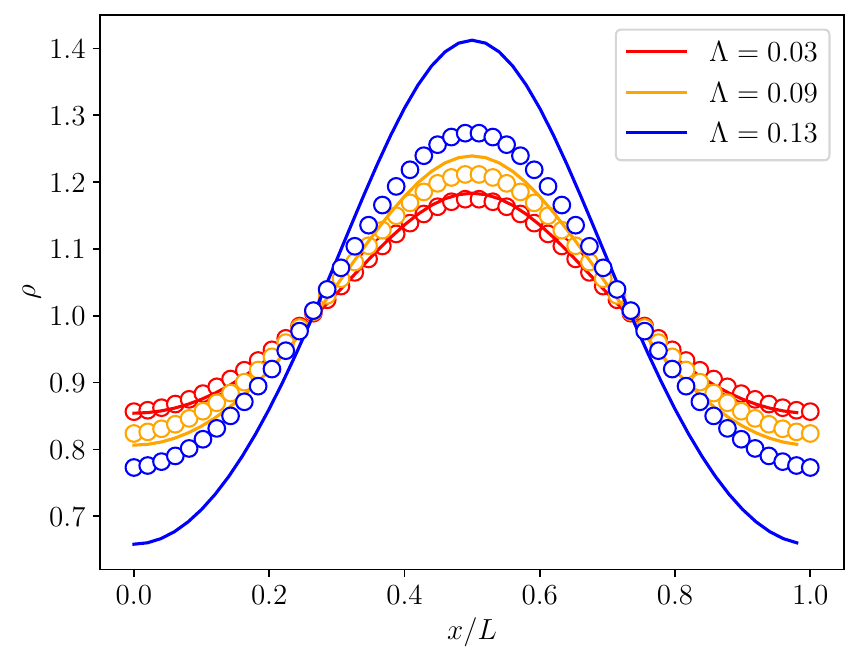}
\caption{Equilibrium density profile $\rho(x)$ for different values of the quantum parameter $\Lambda$ (as in the legend) and with $k_1 = 0.3$. Open circles are the numerical calculations (see Appendix~\ref{app:schr}) and lines are the prediction of Eq.~\eqref{eq:rho_eq}. 
\label{fig:rho}}
\end{figure}

\begin{figure}
\includegraphics[scale=0.5]{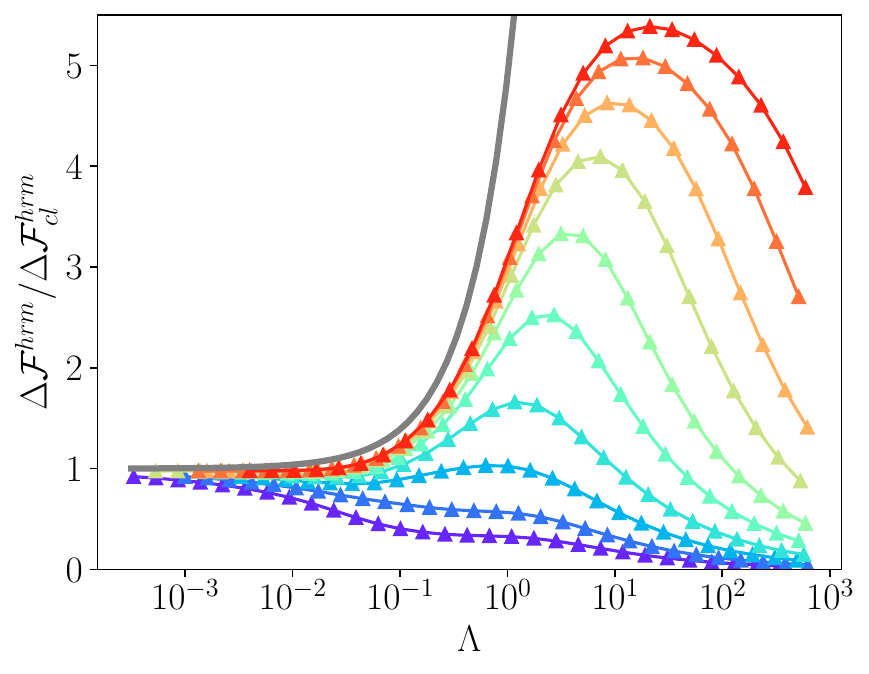}
\caption{Free energy barrier as a function of $\Lambda$ for the harmonic potential for different values of $L_x/L_\omega = 2.8, 3.7,  5.0, 6.6, 8.9, 11.7, 15.8, 20.8,  28.1, 37.0$ from violet to red. Lines with points are obtained from the numerical solution of the Schr\"odinger equation while the grey line is the prediction of the Fick-Jacobs model.
\label{fig:DF_temp}}
\end{figure}

In order to validate our model, in Fig.~\ref{fig:rho}, we compare the density profile obtained from Eqs.~\eqref{eq:free_en-harm} and~\eqref{eq:rho_eq} against the density profile obtained from the numerical solution of the quantum master equation describing the system coupled to a thermal bath at temperature $T$. 
The quantum master equation is derived by first discretizing the Schr\"odinger equation (see Appendix~\ref{app:schr})
\begin{align}
i\hbar \partial_t \psi(x,y,t) = \left[-\frac{\hbar^2}{2m} (\partial_x^2+\partial_y^2)+ U(x,y) \right] \psi(x,y,t),
\end{align}
for the system's wave function $\psi(x,y,t)$, by introducing a set of grid points $(x_i,y_j) = (\Delta_x (i-1/2),\Delta_y j)$ with spacing $\Delta_x$ and $\Delta_y$ and corresponding field operators
\begin{align}
\hat{a}_{i,j} = (\Delta_x \Delta_y)^{-1/2} \int\limits_{x_i-\Delta_x/2}^{x_i+\Delta_x/2} \mathrm{d}x\int\limits_{y_j-\Delta_y/2}^{y_j+\Delta_y/2} \mathrm{d}y \, \hat{\psi}(x,y)\,. 
\end{align}
Here, $\hat{\psi}(x,y)$ is the field operator of the underlying continuous bosonic field. 
Note that we transform the Schrödinger equation to second-quantized form to later allow for particle exchange with a bath.
As discussed in  
Appendix~\ref{app:schr}, the system Hamiltonian then becomes
\begin{align}
\begin{split}
\hat{H}_\mathrm{S} =\sum_{ij} &\biggl[-J_x \left( \hat{a}_{i+1,j}^\dagger \hat{a}_{i,j} + \hat{a}_{i,j}^\dagger \hat{a}_{i+1,j} - 2 \hat{a}_{i,j}^\dagger \hat{a}_{i,j}\right)
\\
&-J_y \left( \hat{a}_{i,j+1}^\dagger \hat{a}_{i,j} + \hat{a}_{i,j}^\dagger \hat{a}_{i,j+1} - 2 \hat{a}_{i,j}^\dagger \hat{a}_{i,j}\right)\\
&+ U(x_i, y_j) \, \hat{a}_{i,j}^\dagger \hat{a}_{i,j}\biggr],
\end{split}
\end{align}
with $J_x = {\hbar^2}/({2m \Delta_x^2})$ and $J_y = {\hbar^2}/({2m \Delta_y^2})$.
We then compute the thermal density matrix $\hat\varrho=\exp(-\beta\hat{H}_\mathrm{S}) / Z$ with partition function $Z=\mathrm{tr}\exp(-\beta\hat{H}_\mathrm{S})$. In Fig.~\ref{fig:rho}, we then show the spatial density in the $x$-direction 
\begin{align}
\varrho(x) = \int_{-\infty}^\infty\! \mathrm{d}y\ \mathrm{tr}\left[ \hat{\psi}^\dagger(x,y)\hat{\psi}(x,y)\hat\varrho\right]\,.    
\end{align}
For the discretized case, this corresponds to  
\begin{align}
\varrho(x_i) = \Delta_x^{-1}\sum_j \mathrm{tr}( \hat{a}_{i,j}^\dagger \hat{a}_{i,j}\hat\varrho)\,.    
\end{align}
Fig.~\ref{fig:rho} shows that the predictions of our analytical model are reliable up to $\Lambda\simeq 0.1$.
Interestingly, in agreement with Eq.~\eqref{eq:rho-lambda}, Fig.~\ref{fig:rho} shows that the net effect of the first quantum corrections is to reduce the density where the local potential barrier is stiffer, at $x/L=0,1$ (i.e., larger values of $k(x)$), and enhance it where $k(x)$ is smaller, at $x/L=1/2$.
In order to further assess the reliability of the model on a wider range of parameters, we focus on the free energy barrier, i.e.~the difference between the free energy maximum (at $x=0$ and $x=L$ in Fig.~\ref{fig:rho}) and minimum (at $x=L/2$ in Fig.~\ref{fig:rho}),
\begin{align}
\beta \Delta \mathcal{F}^{hrm}
=\frac{1}{2}\ln\frac{1+k_{1}}{1-k_{1}}+ 4\Lambda k_1.
\label{eq:DF-lambda}
\end{align}
Surprisingly, Eq.~\eqref{eq:DF-lambda} shows that the net effect of the quantum corrections is to increase the effective free energy barrier. This is counterintuitive since, a priori one would have expected a reduction of the free energy barrier due to ``tunnel" effects. In contrast, Fig.~\ref{fig:DF_temp} shows a kind of ``anti-tunnel" effect, where the free energy barrier increases upon reducing the temperature (and hence enhances the relevance of the quantum corrections). 

To better understand the physical origin of the growth of the free energy barrier upon increasing $\Lambda$ it is interesting to compare Eq.~\eqref{eq:free_en-harm} to the case of a genuine $1\mathrm{D}$ system in a periodic potential. In this case, at linear order in $\Lambda$, the effective potential only has an ``enthalpic" contributions and it reads
\begin{align}
\psi_{ent}(x) = U(x)+ \frac{\lambda^2_T}{48 \pi} \partial_x^2  U(x)\,.
\end{align}
Hence, for the case of 
\begin{align}
U(x) = U_0\cos(2\pi x/L),
\end{align} 
we have that 
\begin{align}
\Delta \psi_{ent} \equiv \psi(0)-\psi(L/2) = 2 U_0\left(1-\frac{\pi}{12}\frac{\lambda^2_T}{L^2}\right)
\end{align}
and hence, as expected, the quantum corrections reduce the effective potential barrier.  
Therefore the enhancement of the free energy barrier in Eq.~\eqref{eq:DF-lambda} is indeed due to the mixed enthalpic-entropic nature of the effective free energy obtained via integrating along the transverse direction.\\ 

To test the reliability of the model, in Fig.~\ref{fig:DF_temp} we compare Eq.~\eqref{eq:DF-lambda} to the free energy barrier obtained from the numerical solution of the Schr\"odinger equation coupled to a thermal bath. 
Fig.~\ref{fig:DF_temp} shows a very good agreement for $\Lambda \lesssim 0.5$ and for $L_x \geq L_\omega$ with quantum harmonic oscillator length scale $L_\omega = 2\sqrt{\hbar/(m \omega)}$ and $\omega^2 = k_0/m$. Moreover, Fig.~\ref{fig:DF_temp} shows a non-monotonic dependence of $\Delta \mathcal{F}$ on $\Lambda$ which cannot be captured by our expansion up to linear order in $\Lambda$. Such non-monotonic dependence reconciliates the anti-tunneling at finite temperature with the eventual tunneling effect at very small temperature (as shown in Fig.~\ref{fig:DF_temp} at very large values of $\Lambda$).


\begin{figure}
\includegraphics[scale = 0.5]{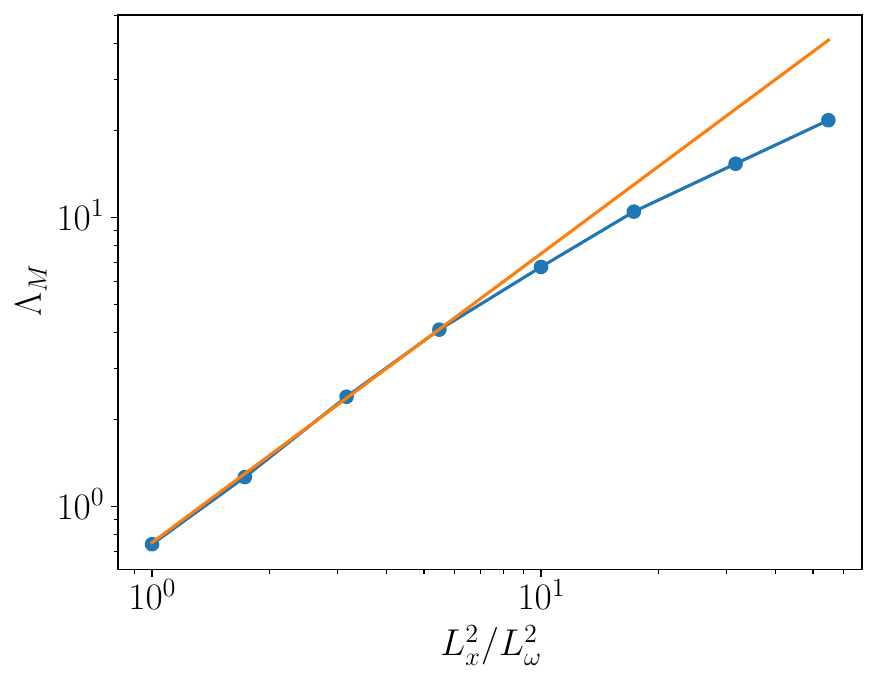}
\caption{Values of $\Lambda$ for which $\Delta F$ attains its maximum, $\Lambda_M$ as a function of $L_x^2/L_\omega^2\propto \sqrt{k_0}$. 
\label{fig:DF_detail}}
\end{figure}
In order to discuss the dependence of the position of the maximum of the free energy barrier $\Delta F_M$ on $\Lambda$, we have performed more refined numerical solutions (see Appendix~\ref{app:DF_max}). In the upper panel of  Fig.~\ref{fig:DF_detail}, we show the location of the maxima $\Lambda_M$ as a function of the relative confinement parameter $L_x/L_\omega$. Interestingly, $\Lambda_M$ initially scales as 
\begin{align}
    \Lambda_M \sim \sqrt{k_0} \propto \frac{L_x^2}{L_\omega^2},
\end{align}
which, apart of prefactors, can be read as the following ratio between length scales
\begin{align}
    \frac{\lambda^2_T}{L_y^2}\simeq \frac{L_x^2}{L_\omega^2}.
\end{align}
Using the definition of $L_y$ and $L_\omega$ it is possible to derive the relation 
$
    L^2_\omega \simeq \lambda_T L_y
$ 
and hence, by rearranging terms, the maximum occurs for 
\begin{align}
    \frac{\lambda^2_T}{L_x^2}\simeq \frac{L_y}{\lambda_T}.
    \label{eq:max}
\end{align}
Equation~\eqref{eq:max} shows that when the length scale separation along the $x$-axis is stronger than the one on the $y$-axis i.e., for $\lambda^2_T/L_x^2 \ll L_y/\lambda_T$ the first order corrections to the Smoluchowski equation are the leading one and hence the analytical model follows the numerical results: the free energy grows upon increasing $\Lambda$ hence leading to \textit{anti-tunneling}. In contrast, for $\lambda^2_T/L_x^2 \gg L_y/\lambda_T$ the higher-order corrections become important and the effective free energy barrier decreases, retrieving, for $\Lambda \rightarrow \infty$ the usual \textit{tunneling} effect. 
\begin{figure}
    \centering
    \includegraphics[scale=0.5]{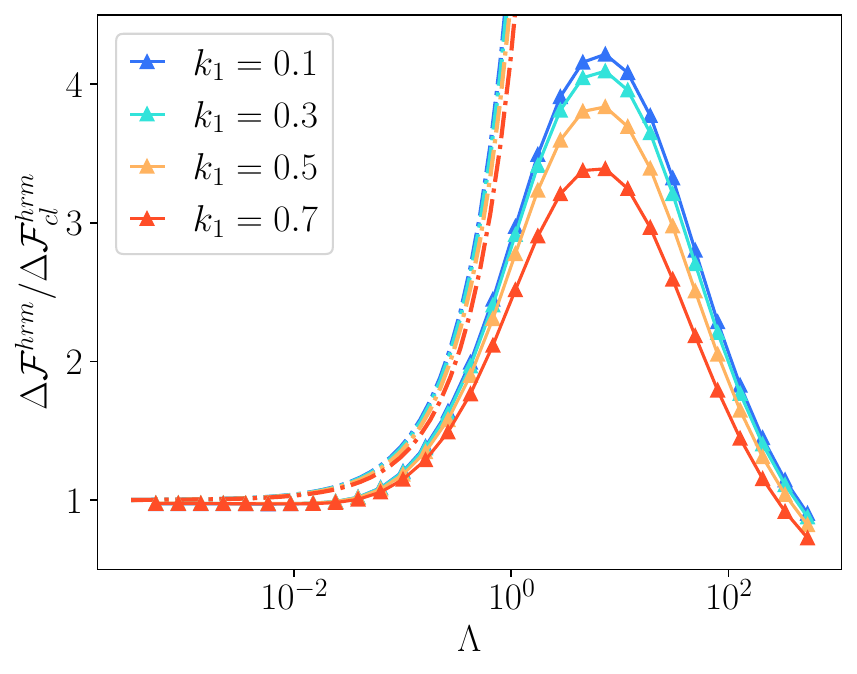}
    \caption{Free energy barrier as a function of $\Lambda$ for the harmonic potential for different values of $k_1$ (see legend) and for $L_x/L_\omega = 15.8$. Lines with points are obtained from the numerical solution of the Schr\"odinger equation while dot-dashed lines are the prediction of the Fick-Jacobs model.}
    \label{fig:DF_VS_K1}
\end{figure}
Finally, we remark that, once the maximum is present, its location  is independent of the amplitude of the corrugation of the potential, $k_1$, as shown in Fig.~\ref{fig:DF_VS_K1}. This supports the fact that $k_1$ does not appear in the scaling relation in Eq.~\eqref{eq:max}. 

\section{Transport}

\begin{figure}
\includegraphics[scale=0.5]{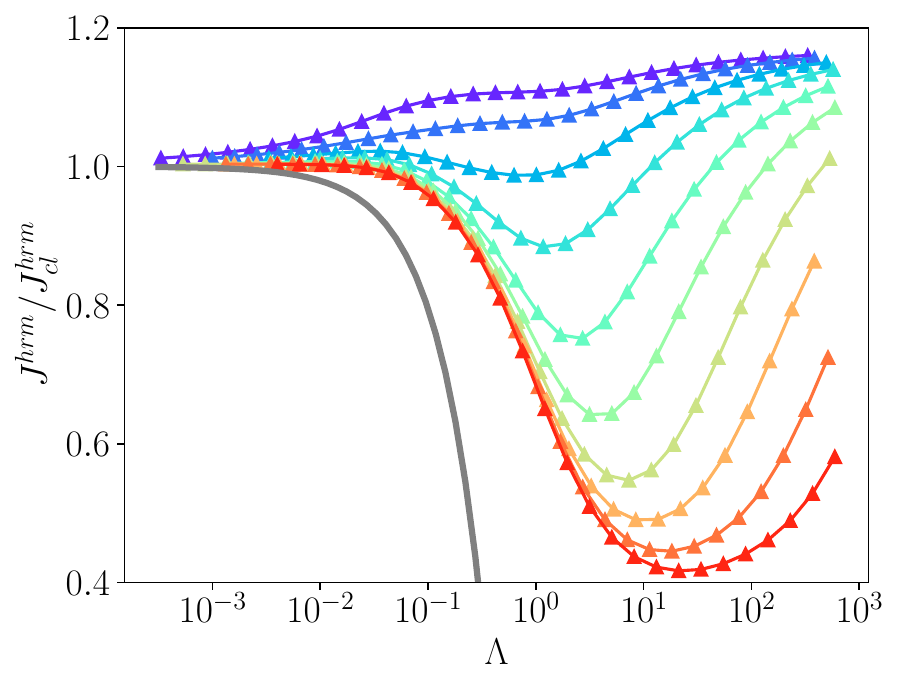}
\caption{Flux as a function of $\Lambda$ for the harmonic potential for different values of $L_x/L_\omega = 2.8, 3.7,  5.0, 6.6, 8.9, 11.7, 15.8, 20.8,  28.1, 37.0$ from violet to red. Lines with points are obtained from the numerical solution of the Schr\"odinger equation while the grey line is the prediction of the Fick-Jacobs model.
\label{fig:flux}}
\end{figure}

In the following, we fix the magnitude of the chemical potential, $\mu_{1,2}$, at the ends (located at $x_{1}=0$, $x_2 =L$) of a period of the external potential. 
Accordingly, for the particle density we have
\begin{align}
    p(x_{1,2},y) = e^{-\beta\psi(x_{1,2},y)}e^{\beta\mu_{1,2}} 
\end{align}
and for the $1\mathrm{D}$ projected density, at linear order in $\Lambda$, we have 
\begin{align}
    \rho_{1,2} &= \int e^{-\beta\psi(x_{1,2},y)}e^{\beta\mu_{1,2}} dy\\
    & \simeq e^{\beta\mu_{1,2}} \!\!\int \!e^{-\beta U}\left[1-\Lambda L_y^2\left(2\partial_y^{2}\beta U-(\partial_{y}\beta U)^{2}\right)\right]dy\nonumber.
\end{align}
Accordingly, we have 
\begin{subequations}
\begin{align}
\Pi & =\rho_{1} e^{\beta \mathcal{F}(x_1)}\simeq \rho_1 e^{\beta A_0(x_1)}\left(1+\Lambda \mathcal{F}(x_1)\right),\\
\frac{J}{D_{cl}} & =\dfrac{\rho_1 e^{\beta \mathcal{F}(x_1)}-\rho_{2}e^{\beta \mathcal{F}(x_2)}}{\int_{x_1}^{x_2}e^{\beta \mathcal{F}(x)}dx}.
\end{align}
\end{subequations}
In the case of  periodic potentials,  for which we have $A_{0}(x_2=L)=A_{0}(x_1=0)$, and at linear order in $\Lambda$ the last expression simplifies to 
\begin{align}
\frac{J}{D_{cl}}  \simeq \frac{J_{cl}}{D_{cl}}\left[1+\Lambda J_\Lambda\right],
\end{align}
with
\begin{subequations}
\begin{align}
 \frac{J_{cl}}{D_{cl}} & = \dfrac{e^{\beta \mu_{1}}-e^{\beta \mu_{2}}}{\int_{0}^{L}e^{\beta A_{0}(x)}dx},\\
 J_\Lambda &= -\dfrac{\int_{0}^{L}e^{\beta A_{0}(x)}\beta \mathcal{F}_\Lambda(x) dx}{\int_{0}^{L}e^{\beta A_{0}(x) }dx}.
\end{align}
\end{subequations}
Finally, for a harmonic potential, this reduces to 
\begin{subequations}
\begin{align}
\frac{J_{cl}^{harm}}{D_{cl}} & = \dfrac{e^{\beta \mu_{1}}-e^{\beta \mu_{2}}}{\int_{0}^{L}\sqrt{{k(x)}/{k_0}}dx}, \\
J_\Lambda & = - 2\dfrac{\int_{0}^{L}\left({k(x)}/{k_0}\right)^\frac{3}{2} dx}{\int_{0}^{L}\sqrt{{k(x)}/{k_0}} dx}.
\label{eq:J-Lambda}
\end{align}
\end{subequations}
We remark that the dependency of the quantum correction on the magnitude of the potential, encoded in $k_0$, is only in $\mathcal{F}_\Lambda(0)$ whereas, recalling the definition of $k(x)$ in Eq.~\eqref{eq:pot},  $J_\Lambda$ is independent of $k_0$. 
Fig.~\ref{fig:flux} shows that the net flux decreases upon increasing $\Lambda$. This is in agreement with the increase in the free energy barrier shown in Fig.~\ref{fig:DF_temp}.
It is interesting to note that even for $k_1=0$ there is still a correction to the current due to the fact that the densities of particles at the ends of the potential are enhanced by the quantum corrections (see Eq.~\eqref{eq:corr-rho-lambda}). In order to assess the validity of Eq.~\eqref{eq:J-Lambda} numerically, we 
couple the system to two particle leads, with gives the total Hamiltonian of system and baths
\begin{align}
\hat{H}_\mathrm{S} = \hat{H}_\mathrm{S} + \sum_{j} \left(\hat{H}^{(j)}_{\mathrm{SB},l}+ \hat{H}^{(j)}_{\mathrm{SB},r} + \hat{H}^{(j)}_{\mathrm{B},l}+ \hat{H}^{(j)}_{\mathrm{B},r}\right).
\end{align}
Here enter the bath Hamiltonians 
\begin{align}
\hat{H}^{(j)}_{\mathrm{B},l} = \sum_k E^\mathrm{B}_k \hat{c}^{\dagger}_{k,j} \hat{c}_{k,j}, \qquad \hat{H}^{(j)}_{\mathrm{B},r} = \sum_k E^\mathrm{B}_k \hat{d}^{\dagger}_{k,j} \hat{d}_{k,j},
\end{align}
with bath energies $E^\mathrm{B}_k$ and corresponding bath annihilation operators $\hat{c}_{k,j}, \hat{d}_{k,j}$. We assume the system-bath Hamiltonian
\begin{align}
\hat{H}^{(j)}_{\mathrm{SB},l} &=  \sum_{k,j} g_{k} \left(\hat{c}^{\dagger}_{k,j} \hat{a}_{1,j} + \hat{a}^{\dagger}_{1,j} \hat{c}_{k,j}  \right), \\
\hat{H}^{(j)}_{\mathrm{SB},r} &=  \sum_{k,j} g_{k} \left(\hat{d}^{\dagger}_{k,j} \hat{a}_{M_x,j} + \hat{a}^{\dagger}_{M_x,j} \hat{d}_{k,j}  \right),
\end{align}
with coupling coefficients $g_{k}$ to bath mode $k$. We choose a constant spectral density
$J(E) = \sum_k g_k^2 \delta(E-E^\mathrm{B}_k)= \gamma$.
As we discuss in Appendix~\ref{App:A2}, we perform the standard Born- and Markov approximation to arrive at a Redfield master equation \cite{BreuerPetruccioneBook} for the density matrix of the reduced system. From this, we derive (see Appendix~\ref{App:A2}) a kinetic equation for the single particle density matrix
$\sigma_{ij,kl}=\mathrm{tr}( \hat{a}_{k,l}^\dagger \hat{a}_{i,j}\hat\varrho)$. This allows us to numerically extract the particle current into the bath in the steady state.

\begin{figure}
\includegraphics[scale = 0.5]{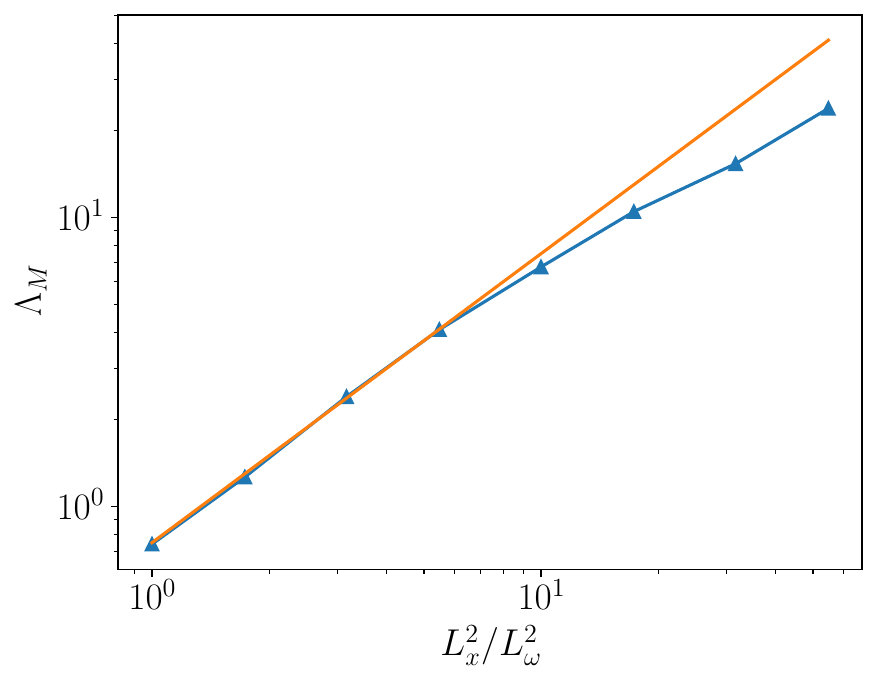}
\caption{Values of $\Lambda$  for which $J$ attains its minimum, $\Lambda_M$, as a function of $L_x^2/L_\omega^2\propto \sqrt{k_0}$. 
\label{fig:flux_detail}}
\end{figure}

Interestingly, the non-monotonic dependence of $\Delta \mathcal{F}$ reported in Fig.~\ref{fig:DF_temp} is mapped into the non-monotonic dependence of $J_\Lambda$ on $\Lambda$ in Fig.~\ref{fig:flux}.
In particular, similarly to what happens for $\Delta \mathcal{F}$, the analytical model (which is linear in $\Lambda$) only captures the reduction in the flux and cannot capture the increase at larger values of $\Lambda$. The dependence of the location of the minimum of the flux follows a similar trend to that of the maximum of $\Delta \mathcal{F}$, as shown in Fig.\ref{fig:flux_detail}.

\section{Discussion}
In order to quantify the magnitude of $\Lambda$, we have summarized in Table~\ref{tab:table} the range of parameters in typical experimental situations in quantum gas experiments similar to Ref.~\cite{husmann_connecting_2015}.   
\begin{table}
\centering
\begin{equation*}
\begin{array}{cclcrcl}
\hline
T &=&\bar{T}\cdot 10^{-6}K & \rightarrow & \bar{T}  &\in & [10^{-2}:\infty]\\
\Delta V &=&\Delta \bar{V} \cdot k_B\cdot10^{-6}\text{K} & \rightarrow & \Delta \bar{V} &\in& [1:100]\\
\lambda &=& \bar{\lambda}\cdot 10^{-6}\text{m} & \rightarrow & \bar{\lambda} &\in& [1:10]\\
\lambda^2_T &=& 3 \bar{\lambda}^2_T\cdot 10^{-12}\text{m}^2 & \rightarrow & \bar{\lambda}^2_T &\in& [0.01:7]\\
\hline
\end{array}
\end{equation*}
\caption{Range of typical experimental values in quantum gas experiments. $\Delta V$ is the depth of the optical potential and  $\lambda$ its wavelength, $\lambda_T$ is the thermal wavelength with $\bar{\lambda}_T=1/(\bar{m}\bar{T})$, and $\bar{m}$ the mass of the particle in atomic units.}
\label{tab:table}
\end{table}
In addition to the definitions in Tab.~\ref{tab:table}, we define $k_0=\Delta V/\lambda^2$. Accordingly, we get
\begin{align}
\beta k_0 \lambda_T^2 \simeq 3 \frac{\Delta \bar{V} \bar{\lambda}^2_T}{\bar{T}\bar{\lambda}^2}=3 \frac{\Delta \bar{V} }{\bar{T}^2\bar{m}\bar{\lambda}^2}
\end{align}
From Tab.~\ref{tab:table} the range of values of $\lambda_T$ is
\begin{align}
\beta k_0 \lambda_T^2 \in [0:10^3],
\end{align}
which correponds to achievable values of 
\begin{align}
\Lambda \in [0:10],
\end{align}
Optical potentials are highly tunable experimentally and should rather easily allow for the trap geometries discussed in this paper. Hence, from this discussion, we expect that our findings should be observable  in transport experiments similar to Ref.~\cite{husmann_connecting_2015} with noninteracting fermions or bosons at low densities (where effects of quantum statistics can be ignored).

We remark that our expansion in $\Lambda$ requested that $\Lambda L_y^2 \partial^2_y U\ll 1$, which implies 
\begin{align}
\beta k_0 \lambda^2_T \ll 48 \pi.
\label{eq:cond-lambda}
\end{align}
Moreover, in our derivation we have required 
\begin{align}
\partial_x U(x,y) \ll \partial_y U(x,y),\\
\partial^2_x U(x,y) \ll \partial^2_y U(x,y).
\end{align}
Accordingly, using Eq.~\eqref{eq:pot}, we have
\begin{align}
\beta k_0 L^2 \gg \frac{4k_1\pi}{1-k_1}.
\label{eq:k0-cond}
\end{align}
Since the typical length $L$ fairly exceeds the thermal wavelength, the latter condition is always fulfilled once Eq.~\eqref{eq:cond-lambda} is fulfilled.

\section{Conclusions}
We have studied the equilibrium and transport properties of a quantum particle in a $2\mathrm{D}$  harmonic confinement potential. We have extended the Fick-Jakobs approximation to the case of the Quantum Smoluchowski equations and identified the effective free energy profile along the longitudinal direction. Interestingly, in the presence of a length scale separation between the longitudinal and transverse direction, our analytical results show that the effective free energy barrier $\Delta\mathcal{F}$ \emph{grows} upon decreasing the temperature, i.e., upon increasing the quantum corrections (up to linear order in $\Lambda$). We thus observe an anti-tunneling phenomenon, which contrasts with the intuitive expectation that the barrier would monotonically decrease as one enters the quantum regime due to the onset of tunneling effects. The barrier enhancement is absent for purely enthalpic potentials. 
We have supported the analytical derivation by numerical solutions of the quantum master equation describing the open system in contact with thermal baths. Our numerical results not only confirm the analytical predictions but also show that for larger values of $\Lambda$ (i.e., lower temperatures) or in the absence of length scale separation, the reduction of the free energy barrier upon increasing $\Lambda$ is retrieved, consistently with expectations. This holds for both the equilibrim free energy barrier, $\Delta\mathcal{F}$, as well as for the flux of particles in the case in which a single period of the potential is putting in contact reservoirs of particles kept at different chemical potentials. 

In future work, especially with regard to quantum gas experiments, it would be interesting to study how these effects are generalized to the case of a $3\mathrm{D}$ trap geometry. Furthermore, it would be interesting to investigate the effects of quantum statistics and interactions on the anti-tunneling--tunneling transition.

\begin{acknowledgements}
The authors thank Andr\'e Eckardt for useful discussions. This work has been partially supported by the Deutsche Forschungsgemeinschaft (DFG, German Research Foundation) via the Reasearch Unit FOR 5688 (Project No.\ 521530974).
\end{acknowledgements}

\bibliography{quantum_fj.bib,quantum_lit}

\clearpage

\onecolumngrid
\appendix

\section{Microscopic open system description\label{app:schr}}

\subsection{System Hamiltonian}

Let us discuss the quantum dynamics of the isolated system (the 2D channel) first. Here, in first quantization, the dynamics is given by the 2D Schrödinger equation

\begin{align}
i\hbar \partial_t \psi(x,y,t) = \left[-\frac{\hbar^2}{2m} (\partial_x^2+\partial_y^2)+ U(x,y) \right] \psi(x,y,t),
\end{align}
where we assume $0\leq x \leq L$ and $y \in \mathbb{R}$.

Let us first transform this equation to second-quantized form (to later allow for particle exchange with a bath). This gives rise to the second-quantized system Hamiltonian
\begin{align}
\hat{H}_\mathrm{S} =\int_0^L \mathrm{d}x\int_{-\infty}^\infty \mathrm{d}y \, \hat{\psi}^\dagger(x,y) \left[-\frac{\hbar^2}{2m} (\partial_x^2+\partial_y^2)+ U(x,y) \right] \hat{\psi}(x,y),
\end{align}
with field operator $\hat{\psi}(x,y)$. Since we want to describe single-particle physics, the quantum statistical properties of the particle field are irrelevant, so we choose a bosonic field $\hat{\psi}(x,y)$ for convenience.
In order to solve the dynamics numerically, let us introduce a discrete grid of spatial points
\begin{align}
(x,y) \rightarrow (x_i,y_j) = (\Delta_x (i-1/2),\Delta_y j)
\end{align}
with $\Delta_x = L/M_x$, and $i=1, \dots, M_x$, as well as $\Delta_y=2\tilde L_y /M_y$, with $j=-M_y/2, -M_y/2+1, \dots, M_y/2$ and some cutoff length $\tilde L_y$ that should be chosen such that the steady state real space density can be neglected for $\vert y \vert > \tilde L_y$.

This allows us to introduce the new field operators on this grid,
\begin{align}
\hat{a}_{i,j} = \frac{1}{\sqrt{\Delta_x \Delta_y}}\int_{x_i^-}^{x_i^+} \mathrm{d}x\int_{y_j^-}^{y_j^+} \mathrm{d}y \, \hat{\psi}(x,y),
\end{align}
where we define $x_i^{\pm}=x_i\pm\Delta_x/2$, and  $y_j^{\pm}=y_j\pm\Delta_y/2$.
Note that with this definition we indeed recover discrete bosonic field operators since 
\begin{align}
[\hat{a}_{i,j}, \hat{a}^\dagger_{i',j'}] = \frac{1}{\Delta_x \Delta_y}\int_{x_i^-}^{x_i^+} \mathrm{d}x\int_{y_j^-}^{y_j^+} \mathrm{d}y \int_{x_{i'}^-}^{x_{i'}^+} \mathrm{d}x'\int_{y_{j'}^-}^{y_{j'}^+} \mathrm{d}y' \, [\hat{\psi}(x,y), \hat{\psi}^\dagger(x',y')] = \delta_{ii'}\delta_{jj'}.
\end{align}
Finally, by replacing the second order derivative with the correponding discrete operator, we find
\begin{align}
\hat{H}_\mathrm{S} =\sum_{ij} \biggl[-&J_x \left( \hat{a}_{i+1,j}^\dagger \hat{a}_{i,j} + \hat{a}_{i,j}^\dagger \hat{a}_{i+1,j} - 2 \hat{a}_{i,j}^\dagger \hat{a}_{i,j}\right)-J_y \left( \hat{a}_{i,j+1}^\dagger \hat{a}_{i,j} + \hat{a}_{i,j}^\dagger \hat{a}_{i,j+1} - 2 \hat{a}_{i,j}^\dagger \hat{a}_{i,j}\right) + U(x_i, y_j) \, \hat{a}_{i,j}^\dagger \hat{a}_{i,j}\biggr],
\label{eq:sys-hamil-disc}
\end{align}
with $J_x = {\hbar^2}/({2m \Delta_x^2})$ and $J_y = {\hbar^2}/({2m \Delta_y^2})$.

\subsection{Coupling to thermal particle reservoirs\label{App:A2}}
We now couple to the system to two thermal reservoirs with Temperature $T_{l/r}$ and chemical potential $\mu_{l/r}$ on each of the leftmost and the rightmost sites. This leads to the total Hamiltonian of system and baths
\begin{align}
\hat{H}_\mathrm{S} = \hat{H}_\mathrm{S} + \sum_{j} \left(\hat{H}^{(j)}_{\mathrm{SB},l}+ \hat{H}^{(j)}_{\mathrm{SB},r} + \hat{H}^{(j)}_{\mathrm{B},l}+ \hat{H}^{(j)}_{\mathrm{B},r}\right).
\end{align}
Here enter the bath Hamiltonians 
\begin{align}
\hat{H}^{(j)}_{\mathrm{B},l} = \sum_k E^\mathrm{B}_k \hat{c}^{\dagger}_{k,j} \hat{c}_{k,j}, \qquad \hat{H}^{(j)}_{\mathrm{B},r} = \sum_k E^\mathrm{B}_k \hat{d}^{\dagger}_{k,j} \hat{d}_{k,j},
\end{align}
with bath energies $E^\mathrm{B}_k$ and corresponding bosonic bath annihilation operators $\hat{c}_{k,j}, \hat{d}_{k,j}$. We assume the system-bath Hamiltonian
\begin{align}
\hat{H}^{(j)}_{\mathrm{SB},l} =  \sum_{k,j} g_{k} \left(\hat{c}^{\dagger}_{k,j} \hat{a}_{1,j} + \hat{a}^{\dagger}_{1,j} \hat{c}_{k,j}  \right), \qquad \hat{H}^{(j)}_{\mathrm{SB},r} =  \sum_{k,j} g_{k} \left(\hat{d}^{\dagger}_{k,j} \hat{a}_{M_x,j} + \hat{a}^{\dagger}_{M_x,j} \hat{d}_{k,j}  \right),
\end{align}
with coupling coefficients $g_{k}$ to mode $k$ to the left bath or to the right bath, respectively.

Using standard Born- and Markov approximation \cite{BreuerPetruccioneBook}, we find the Redfield master equation for the reduced density matrix of the system,
\begin{align}
\partial_t \hat{\varrho}(t) = -i [\hat{H}_\mathrm{S} , \hat{\varrho}(t)] &+ \sum_j \left( [\hat{\tilde{a}}^\dagger_{1,j} \hat{\varrho}(t), \hat{a}_{1,j}] + \mathrm{h.c.}\right) + \sum_j \left( [\hat{\tilde{a}}_{1,j} \hat{\varrho}(t), \hat{a}^\dagger_{1,j}] + \mathrm{h.c.}\right) \\
& + \sum_j \left( [\hat{\tilde{a}}^\dagger_{M_x,j} \hat{\varrho}(t), \hat{a}_{M_x,j}] + \mathrm{h.c.}\right) + \sum_j \left( [\hat{\tilde{a}}_{M_x,j} \hat{\varrho}(t), \hat{a}^\dagger_{M_x,j}] + \mathrm{h.c.}\right).
\end{align}
Herein occur the jump operators
\begin{align}
\hat{\tilde{a}}^\dagger_{1,j} &= \int_0^\infty \mathrm{d}\tau\,  e^{-\frac{i}{\hbar} \hat{H}_\mathrm{S} \tau}  \hat{a}^\dagger_{1,j} e^{\frac{i}{\hbar} \hat{H}_\mathrm{S} \tau} \sum_k g_k^2\, \langle e^{\frac{i}{\hbar} \hat{H}_\mathrm{B} \tau}  \hat{c}^\dagger_{k,j} e^{-\frac{i}{\hbar} \hat{H}_\mathrm{B} \tau} \hat{c}_{k,j} \rangle
\intertext{and}
\hat{\tilde{a}}_{1,j} &= \int_0^\infty \mathrm{d}\tau\,  e^{-\frac{i}{\hbar} \hat{H}_\mathrm{S} \tau}  \hat{a}_{1,j} e^{\frac{i}{\hbar} \hat{H}_\mathrm{S} \tau} \sum_k g_k^2\, \langle e^{\frac{i}{\hbar} \hat{H}_\mathrm{B} \tau}  \hat{c}_{k,j} e^{-\frac{i}{\hbar} \hat{H}_\mathrm{B} \tau} \hat{c}^\dagger_{k,j} \rangle.
\end{align}
Analogous expressions hold for the jump operators on the right hand side of the system. Here we calculate 
\begin{align}
\langle e^{\frac{i}{\hbar} \hat{H}_\mathrm{B} \tau}  \hat{c}^\dagger_{k,j} e^{-\frac{i}{\hbar} \hat{H}_\mathrm{B} \tau} \hat{c}_{k,j} \rangle = e^{\frac{i}{\hbar}  E^\mathrm{B}_k \tau} n_l(E^\mathrm{B}_k) 
\end{align}
as well as
\begin{align}
\langle e^{\frac{i}{\hbar} \hat{H}_\mathrm{B} \tau}  \hat{c}_{k,j} e^{-\frac{i}{\hbar} \hat{H}_\mathrm{B} \tau} \hat{c}^\dagger_{k,j} \rangle = e^{-\frac{i}{\hbar} E^\mathrm{B}_k \tau} (1+n_l(E^\mathrm{B}_k))\overset{n_l(E) \ll 1}{\longrightarrow} e^{-\frac{i}{\hbar} E^\mathrm{B}_k \tau},
\end{align}
and in the second expression we have used the classical limit in which the population of individual modes is very small, $n_l(E) \ll 1$.
Herein occurs the Bose-Einstein occupation function for the left bath, where we again perform the classical limit
\begin{align}
n_l(E) = \frac{1}{e^{\beta (E-\mu_l)}-1} \overset{n_l(E) \ll 1}{\longrightarrow} e^{-\beta (E-\mu_l)} = n_{\mathrm{cl},l}(E).
\end{align}
We introduce the spectral density of the baths (we assume equal spectral densities for both baths)
\begin{align}
J(E) = \sum_k g_k^2 \delta(E-E^\mathrm{B}_k)= \gamma,
\end{align}
which we choose to be constant.
It also enter the single-particle eigenstates of the system $\varphi_k(i,j)$ and corresponding system eigenenergies $E_k$, such that the field operators
\begin{align}
\hat b_k = \sum_{i,j} \varphi^*_k(i,j) \hat a_{i,j}, \qquad \hat b^\dagger_k = \sum_{i,j} \varphi_k(i,j) \hat a^\dagger_{i,j}
\end{align}
diagonalize the system Hamiltonian
\begin{align}
\hat{H}_\mathrm{S} = \sum_k E_k \hat b^\dagger_k \hat b_k. 
\end{align}
With this, we have
\begin{align}
e^{-\frac{i}{\hbar} \hat{H}_\mathrm{S} \tau}  \hat{a}^\dagger_{1,j} e^{\frac{i}{\hbar} \hat{H}_\mathrm{S} \tau} = \sum_k \hat b^\dagger_k \varphi^*_k(1, j) e^{-\frac{i}{\hbar} E_k \tau} = \sum_{i,l} \hat a^\dagger_{i,l} \sum_k \varphi_k(i,l) \varphi^*_k(1, j) e^{-\frac{i}{\hbar} E_k \tau},
\end{align}
and 
\begin{align}
e^{-\frac{i}{\hbar} \hat{H}_\mathrm{S} \tau}  \hat{a}_{1,j} e^{\frac{i}{\hbar} \hat{H}_\mathrm{S} \tau} = \sum_k \hat b_k \varphi_k(1, j) e^{\frac{i}{\hbar} E_k \tau} = \sum_{i,l} \hat a_{i,l} \sum_k \varphi^*_k(i,l) \varphi_k(1, j) e^{\frac{i}{\hbar} E_k \tau}.
\end{align}
By using the Sokhotski–Plemelj theorem and neglecting the imaginaries of the Fourier transform, we finally have
\begin{align}
\hat{\tilde{a}}^\dagger_{1,j} &=  \sum_{i,l} \hat a^\dagger_{i,l} f^{i,l}_{1,j},\qquad \hat{\tilde{a}}_{1,j} =  \sum_{i,l} \hat a_{i,l} g^{i,l}_{1,j}.
\end{align}
Herein we define the coefficients
\begin{align}
f^{i,l}_{m,j} &= \sum_k \varphi_k(i,l) \varphi^*_k(m, j) J(E_k) n_{\mathrm{cl}}(E_k),\\
g^{i,l}_{m,j} &= \sum_k \varphi^*_k(i,l) \varphi_k(m, j) J(-E_k). 
\end{align}
This gives rise to the master equation in  explicit form with respect to local creation and annihilation operators, reading
\begin{align}
\partial_t \hat{\varrho}(t) = -i [\hat{H}_\mathrm{S} , \hat{\varrho}(t)] &+ \sum_{ilj} \left( f^{i,l}_{1,j} [\hat{a}^\dagger_{i,l} \hat{\varrho}(t), \hat{a}_{1,j}] + \mathrm{h.c.}\right) + \sum_{ilj} g^{i,l}_{1,j} \left( [\hat{a}_{i,l} \hat{\varrho}(t), \hat{a}^\dagger_{1,j}] + \mathrm{h.c.}\right) \\
& + \sum_{ilj} \left( f^{i,l}_{M_x,j} [\hat{a}^\dagger_{i,l} \hat{\varrho}(t), \hat{a}_{M_x,j}] + \mathrm{h.c.}\right) + \sum_{ilj} g^{i,l}_{M_x,j} \left( [\hat{a}_{i,l} \hat{\varrho}(t), \hat{a}^\dagger_{M_x,j}] + \mathrm{h.c.}\right).
\end{align}

From this master equation, we can derive a kinetic equation for the single-particle density matrix
\begin{align}
\sigma_{ij,kl}=\mathrm{tr}( \hat{a}_{k,l}^\dagger \hat{a}_{i,j}\hat\varrho),
\end{align}
which in the continuum limit leads to the distribution $p(x,y)$ of the main text via
\begin{align}
p(x_i, y_i) = \frac{\sigma_{ij,kl}}{\Delta_x \Delta_y}
\end{align}

By rewriting Eq.~\eqref{eq:sys-hamil-disc} with single particle matrix elements, 
\begin{align}
\hat{H}_\mathrm{S} = \sum_{ij,kl} h_{ij,kl} \hat{a}_{i,j}^\dagger \hat{a}_{k,l}
\end{align}
we finally have
\begin{align}
\partial_t \sigma_{ij,kl}=-i&\sum_{nm} \left(h_{ij,nm}\sigma_{nm,kl} - \sigma_{ij,nm}h_{nm,kl} \right) + (\mathcal{L}_l[\sigma])_{ij,kl} + (\mathcal{L}_r[\sigma])_{ij,kl},
\end{align}
with left dissipator
\begin{align}
(\mathcal{L}_l[\sigma])_{ij,kl} =& \delta_{k1}\left[f^{i,j}_{1,l}+ \sum_{nm}   \sigma_{ij,nm} \left(f^{n,m}_{1,l}-(g^{n,m}_{1,l})^* \right) \right] + \delta_{i1}\left[(f^{k,l}_{1,j})^*+ \sum_{nm}  \left((f^{n,m}_{1,j})^*-g^{n,m}_{1,j} \right) \sigma_{nm,kl}\right],
\end{align}
and right dissipator
\begin{align}
(\mathcal{L}_r[\sigma])_{ij,kl} = & \delta_{kM_x}\left[f^{i,j}_{M_x,l}+ \sum_{nm}   \sigma_{ij,nm} \left(f^{n,m}_{M_x,l}-(g^{n,m}_{M_x,l})^* \right) \right] + \delta_{iM_x}\left[(f^{k,l}_{M_x,j})^*+ \sum_{nm}  \left((f^{n,m}_{M_x,j})^*-g^{n,m}_{M_x,j} \right) \sigma_{nm,kl}\right].
\end{align}
We solve this master equation to solve for the steady state
\begin{align}
\partial_t \sigma^\mathrm{SS}_{ij,kl}=0.
\end{align}
The particle current into the left bath
is then given by
\begin{align}
J = \mathrm{tr}[(\partial_t\varrho)_\mathrm{left} N] = \sum_{ij} \mathrm{tr}[(\partial_t\varrho)_\mathrm{left} \hat a^\dagger_{ij} \hat a_{ij}]  = \sum_{ij} (\mathcal{L}_l[\sigma^\mathrm{SS}])_{ij,ij}.
\end{align}
Note that we choose the chemical potentials $\mu_l$ and $\mu_r$ such that the density on the left is and the right is according to $\varrho_1$ and $\varrho_2$ as assumed in the Smoluchowski equation.

\section{Derivation of Eq.~\eqref{eq:F-Lambda}}
\label{sec:app-derivFLamb}

At leading order in $\Lambda$ and for $\partial_{x}\beta U\ll \partial_y U(x,y)$ we have
\begin{align}
e^{-\beta\psi(x,y)} & \simeq e^{-\beta U(x,y)}\left[1-2\Lambda L_y^2 \partial^2_y \beta U+\Lambda L_y^2 \left(\beta\partial_y U\right)^2\right]
\end{align}
we note that 
\begin{equation}
\int_{-\infty}^{\infty}e^{-\beta U(x,y)}\left(\partial_{y}^{2}\beta U(x,y)-\left(\beta\partial_{y}U(x,y)\right)^{2}\right)dy=
-\int_{-\infty}^{\infty}\partial_{y}^{2}\left(e^{-\beta U(x,y)}\right)dy=-\partial_{y}\left[e^{-\beta U(x,y)}\right]_{-\infty}^{\infty}=0
\label{eq:aux1}
\end{equation}
Hence we have
\begin{align}
e^{-\beta A(x)}=\int\limits_{-\infty}^{\infty}e^{-\beta\psi(x,y)}\frac{dy}{L_y} & \simeq
\!\!\int\limits_{-\infty}^{\infty}e^{-\beta U(x,y)}\left[1-\Lambda L_y^2(2\partial^{2}_y\beta U-\left(\beta\partial_y U\right)^{2})\right]\frac{dy}{L_y} \nonumber\\
&\simeq\int_{-\infty}^{\infty}e^{-\beta U}\left[1-\Lambda L_y^2\partial_{y}^{2}\beta U\right]\frac{dy}{L_y}\nonumber\\
& \simeq e^{-\beta A_0(x)}\left[1-\Lambda L_y^2\int_{-\infty}^{\infty}\dfrac{e^{-\beta U}}{e^{-\beta A_0}}\partial_{y}^{2}\beta U\frac{dy}{L_y}\right]
\end{align}
Where we have introduced 
\begin{align}
e^{-\beta A_{0}(x)}=\int_{-\infty}^{\infty}e^{-\beta U(x,y)}\frac{dy}{L_y}
\end{align}
Concerning $\partial_x\mathcal{F}(X)$ we have:
\begin{align}
\beta\partial_x\mathcal{F}(x) =\int\limits_{-\infty}^{\infty}\dfrac{e^{-\beta\psi(x,y)}}{e^{-\beta A(x)}}\beta\partial_{x}U(x,y)\frac{dy}{L_y} & \simeq
\int_{-\infty}^{\infty}\dfrac{e^{-\beta U}}{e^{-\beta A_{0}}}\left[1-2\Lambda L_y^2\partial_{y}^{2}\beta U+\Lambda L_y^2\left(\beta\partial_{y}U\right)^{2}\right]\beta\partial_{x}U\frac{dy}{L_y}+\nonumber\\
 & +\Lambda L_y^2\int_{-\infty}^{\infty}\dfrac{e^{-\beta U}}{e^{-\beta A_{0}}}\beta\partial_{x}U\frac{dy}{L_y}\int_{-\infty}^{\infty}\dfrac{e^{-\beta U}}{e^{-\beta A_{0}}}\left(\partial_{y}^{2}\beta U\right)\frac{dy}{L_y}
\end{align}
Moreover we have 
\begin{equation}
\int\limits_{-\infty}^{\infty}\dfrac{e^{-\beta U(x,y)}}{e^{-\beta A_{0}(x)}}\beta\partial_{x}U(x,y)\frac{dy}{L_y}=
-\int\limits_{-\infty}^{\infty}\dfrac{\partial_{x}\left(e^{-\beta U(x,y)}\right)}{e^{-\beta A_{0}(x)}}\frac{dy}{L_y}=
-\partial_{x}\int\limits_{-\infty}^{\infty}\dfrac{e^{-\beta U(x,y)}}{e^{-\beta A_{0}(x)}}\frac{dy}{L_y}+\int\limits_{-\infty}^{\infty}\dfrac{e^{-\beta U(x,y)}}{e^{-\beta A_{0}(x)}}\frac{dy}{L_y}\beta\partial_{x}A_{0}(x)=\beta\partial_{x}A_{0}(x)
\end{equation}
Finally we have:
\begin{align}
\beta\partial_x\mathcal{F}(x) \simeq
\beta \partial_x A_0(x)\left[1+\Lambda L_y^2\int_{-\infty}^{\infty}\dfrac{e^{-\beta U}}{e^{-\beta A_{0}}}\partial_{y}^{2}\beta U\frac{dy}{L_y}\right]+\int_{-\infty}^{\infty}\dfrac{e^{-\beta U}}{e^{-\beta A_{0}}}\left[-2\Lambda L_y^2\partial_{y}^{2}\beta U+\Lambda L_y^2\left(\beta\partial_{y}U\right)^{2}\right]\beta\partial_{x}U\frac{dy}{L_y}
\end{align}
that can be grouped as 
\begin{align}
\beta\partial_x\mathcal{F}(x)  & \simeq \beta\partial_{x}A_{0}(x)+\Lambda \beta\partial_x\mathcal{F}_\Lambda(x) 
\label{eq:Gamma}
\end{align}
where we have introduced 
\begin{align}
\beta\partial_x\mathcal{F}_\Lambda(x)  =\beta\partial_x A_0(x)L_y^2\int\limits_{-\infty}^{\infty}\dfrac{e^{-\beta U(x,y)}}{e^{-\beta A_{0}(x)}}\partial_{y}^{2}\beta U(x,y)\frac{dy}{L_y}+
L_y^2\int\limits_{-\infty}^{\infty}\dfrac{e^{-\beta U(x,y)}}{e^{-\beta A_{0}(x)}}\left(\left(\beta\partial_{y}U(x,y)\right)^{2}-2\partial_{y}^{2}\beta U(x,y)\right)\beta\partial_{x}U(x,y)\frac{dy}{L_y} 
\end{align}

\section{Derivation of Eq.~\eqref{eq:free_en-harm}\label{app:DF-harm}}
We can further simply the last expression in the case of a harmonic
potential of the form 
\begin{align}
\beta U(x,y) & =\frac{1}{2}\beta k(x)y^{2}\\
k(x) & =k_{0}\left(1+k_{1}\cos\frac{2\pi x}{L}\right)
\end{align}
where $L$ is the period of the potential. 
Recalling that 
\begin{align}
\int_{-\infty}^{\infty}e^{-\frac{1}{2}\beta k(x) y^{2}}\frac{dy}{L_y} & =\sqrt{\frac{2\pi}{\beta k(x) L_y^2}}
\end{align}
For such a potential we have 
\onecolumngrid
\begin{align}
e^{-\beta A_{0}(x)} & =\sqrt{\frac{2\pi}{\beta k(x)L_y^2}}\\
L_y^2 \int_{-\infty}^{\infty}\dfrac{e^{-\beta U}}{e^{-\beta A_{0}}}\partial_{y}^{2}\beta U \frac{dy}{L_y} & =L_y^2\sqrt{\frac{\beta k(x)L_y^2}{2\pi}}\int_{-\infty}^{\infty}e^{-\frac{1}{2}\beta k(x)y^{2}}\beta k(x)\frac{dy}{L_y} =\beta k(x) L_y^2\\
\beta\int\limits\partial_{x}A_{0}(x)L_y^2 \int_{-\infty}^{\infty}\dfrac{e^{-\beta U}}{e^{-\beta A_{0}}}\partial_{y}^{2}\beta U \frac{dy}{L_y}dx & =\frac{1}{2}\beta L_y^2\int\limits\partial_{x}k(x)dx =\frac{1}{2}\beta k(x)L_y^2
\end{align}
and 
\begin{align}
L_y^2\int dx \int\limits_{-\infty}^{\infty}\dfrac{e^{-\beta U(x,y)}}{e^{-\beta A_{0}(x)}}&\left(\left(\beta\partial_{y}U(x,y)\right)^{2}-2\partial_{y}^{2}\beta U(x,y)\right)\beta\partial_{x}U(x,y)\frac{dy}{L_y} =\nonumber\\
& =\frac{1}{2}L_y^2\int\limits\int\limits_{-\infty}^{\infty}\!\!\sqrt{\frac{\beta k(x)L_y^2}{2\pi}}e^{-\frac{1}{2}\beta k(x)y^{2}}\left[\left(\beta k(x)y\right)^{2}-2\beta k(x)\right]\partial_{x}\beta k(x)y^{2}\frac{dy}{L_y}dx\nonumber\\
& = \frac{1}{2}L_y^2\int\limits\sqrt{\frac{\beta k(x)L^2}{2\pi}}\partial_{x}\beta k(x)\left[\beta^2 k^2(x)\int\limits_{-\infty}^{\infty}\!\!e^{-\frac{1}{2}\beta k(x)y^{2}}y^4\frac{dy}{L_y}-2\beta k(x)\int\limits_{-\infty}^{\infty}\!\!e^{-\frac{1}{2}\beta k(x)y^{2}}y^2\frac{dy}{L_y}\right]dx\nonumber\\
&= \frac{1}{2}L_y^2\int\left[3-2\right]\partial_{x}\beta k(x)dx\nonumber\\
&=\frac{1}{2}\beta k(x)L_y^2
\end{align}
Accordingly, we have
\begin{align}
\beta \mathcal{F}(x)&=\beta A_0(x)+\Lambda \beta k(x)L_y^2 = \frac{1}{2}\ln\frac{\beta k(x)L_y^2}{2\pi}+\Lambda \beta k(x) L_y^2
\end{align}
Finally, using $L_y^2=2 k_BT/k_0$ and disregarding terms which do not depend on $x$ we get
\begin{align}
\beta \mathcal{F}(x)&= \frac{1}{2}\ln\frac{k(x)}{k_0}+2\Lambda \frac{k(x)}{k_0}
\label{eq:app-F-harm}
\end{align}
which is Eq.~\eqref{eq:free_en-harm} of the main text.

\section{Harmonic potential: regime of validity}

In our derivation we have requested 
\begin{align}
\partial_x U(x,y) \ll \partial_y U(x,y)\\
\partial^2_x U(x,y) \ll \partial^2_y U(x,y)
\end{align}
This implies:
\begin{align}
y_1\ll L\frac{1+k_1\cos(2\pi x)}{k_1 \pi \sin(2\pi x/L)}\\
y_2\ll L\sqrt{\frac{1+k1\cos(2\pi x)}{2k_1\pi^2\cos(2\pi x/L)}}
\end{align}
In order to fulfill this requirement we impose that when the above mentioned conditions are not fulfilled the probability distribution should be exponentially damped. This implies:
\begin{align}
\frac{1}{2}\beta k y_1\gg 1\\
\frac{1}{2}\beta k y_2\gg 1
\end{align}
Hence we can derive conditions on $k_0$
\begin{align}
\beta k_0 L_y & \gg 2\frac{L_y}{L^2}\frac{\left(k_1 \pi \sin(2 \pi x/L)\right)^2}{\left(1+k_1 \cos(2 \pi x/L)\right)^3}\simeq 2\pi^2\frac{L_y^2}{L^2} k_1^2 \\
\beta k_0 L_y & \gg 4\frac{L_y}{L^2}\frac{k_1 \pi \cos(2 \pi x/L)}{\left(1+k_1 \cos(2 \pi x/L)\right)^2}\simeq 4\pi\frac{L_y^2}{L^2}\frac{k_1\pi}{1-k_1}
\end{align}
whose solution is $\beta k_0 L_y^2 \gg 4\pi L_y^2/L^2k_1/(1-k_1)$.

\section{Transport}
In the following we specialize to the case in which the transport is driven by an imbalance in the chemical potential at the ends of a single period of the potential. Accordingly, we have
\begin{align}
    \rho_{1,2} &= \int e^{-\beta\psi(x_{1,2},y)}e^{\beta\mu_{1,2}} dy\nonumber\\
    & \simeq e^{\beta\mu_{1,2}} \!\!\int \!e^{-\beta U(x_{1,2},y)}\left[1-\Lambda L_y^2\left(2\partial_y^{2}\beta U(x_{1,2},y)-(\partial_{y}\beta U(x_{1,2},y))^{2}\right)\right]dy\nonumber \\
    & \simeq e^{\beta\mu_{1,2}}e^{-\beta A_0(x_{1,2})} \left(1+\Lambda \rho_\Lambda(x_{1,2})\right)\label{eq:app-rho}
\end{align}
with
\begin{align}
    \rho_\Lambda(x_{1,2}) = -\int \dfrac{e^{-\beta U(x_{1,2},y)}}{e^{-\beta A_0(x_{1,2})}} L_y^2\left(2\partial_y^{2}\beta U(x_{1,2},y)-(\partial_{y}\beta U(x_{1,2},y))^{2}\right)dy = -\int \dfrac{e^{-\beta U(x_{1,2},y)}}{e^{-\beta A_0(x_{1,2})}} L_y^2 \partial_y^{2}\beta U(x_{1,2},y) dy
\end{align}
where in the last step we used Eq.~\eqref{eq:aux1}. 
At steady state, the solution of Eq.~\eqref{eq:FJ_qm} reads
\begin{align}
\Pi & =\rho_{1} e^{\beta \mathcal{F}(x_{1})}\simeq e^{\beta\mu_{1}}e^{-\beta A_0(x_{1})} \left(1+\Lambda \rho_\Lambda(x_{1})\right) e^{\beta A_0(x_1)}\left(1+\Lambda \mathcal{F}(x_{1})\right)\simeq e^{\beta\mu_{1,2}}\left(1+\Lambda \rho_\Lambda(x_{1})+\Lambda \mathcal{F}(x_{1})\right)\\
\frac{J}{D_{cl}} 
& =\dfrac{\rho_1 e^{\beta \mathcal{F}(x_{1})}-\rho_{2}e^{\beta \mathcal{F}(x_{2})}}{\int_{x_{1}}^{x_{2}}e^{\beta \mathcal{F}(x)}dx}\nonumber \\
& \simeq\dfrac{e^{\beta \mu_1}\left(1+\Lambda \rho_\Lambda(x_{1})+\Lambda \mathcal{F}(x_{1})\right)-e^{\beta \mu_2}\left(1+\Lambda \rho_\Lambda(x_{2})+\Lambda \mathcal{F}(x_{2})\right)}{\int_{x_{1}}^{x_{2}}e^{\beta A_{0}(x)}\left(1+\Lambda \beta \mathcal{F}_\Lambda(x)\right)dx}
\end{align}
In the case of  periodic potentials for
which we have $A_{0}(x_1)=A_{0}(x_2)$, the last expression simplifies to 
\begin{align}
\frac{J}{D_{cl}}  \simeq \dfrac{e^{\beta \mu_{1}}-e^{\beta \mu_{2}}}{\int_{x_1}^{x_2}e^{\beta A_{0}(x)}dx}
 \left[1+\Lambda \rho_\Lambda(x_{1})+\Lambda\beta \mathcal{F}_\Lambda(x_1)-\Lambda\dfrac{\int_{x_1}^{x_2}e^{\beta A_{0}(x)}\beta \mathcal{F}_\Lambda(x) dx}{\int_{x_1}^{x_2}e^{\beta A_{0}(x) }dx}\right]
\end{align}
and hence the flux can be decomposed in 
$\frac{J}{D_{cl}}\simeq J_{cl}\left[1+\Lambda J_\Lambda\right] $
with 
\begin{align}
    \frac{J_{cl}}{D_{cl}} &= \dfrac{e^{\beta \mu_{1}}-e^{\beta \mu_{2}}}{\int_{x_1}^{x_2}e^{\beta A_{0}(x)}dx}\\
    J_\Lambda &= \rho_\Lambda(x_{1})+\beta \mathcal{F}_\Lambda(x_1)-\dfrac{\int_{x_1}^{x_2}e^{\beta A_{0}(x)}\beta \mathcal{F}_\Lambda(x) dx}{\int_{x_1}^{x_2}e^{\beta A_{0}(x) }dx}
\end{align}

\subsection*{Harmonic potential}
In the case of harmonic potential we have 
\begin{align}
e^{-\beta A_{0}(x)} & =\sqrt{\frac{\pi k_0}{k(x)}}\\
\rho_\Lambda (x_1) & =  -2 \frac{k(x_1)}{k_0}\\
\beta \mathcal{F}_\Lambda (x) &= 2\frac{k(x)}{k_0} 
\end{align}
where we used $L_y^2 = 2k_B T/k_0$. 
Accordingly, we get
\begin{align}
\frac{J_{cl}}{D_{cl}} & = \dfrac{e^{\beta \mu_{1}}-e^{\beta \mu_{2}}}{\int_{x_1}^{x_2}\sqrt{\frac{k(x)}{k_0}}dx} \\
J_\Lambda & = - 2\dfrac{\int_{x_1}^{x_2}\left(\frac{k(x)}{k_0}\right)^\frac{3}{2} dx}{\int_{x_1}^{x_2}\sqrt{\frac{k(x)}{k_0}} dx}
\end{align}

\section{Additional simulations to identify the maximum of $\Delta \mathcal{F}$ and $J$\label{app:DF_max}}
Here we report the additional data that we generated to identify the maxima of $\Delta \mathcal{F}$ and $J$
\begin{figure}[h!]
    \centering
    \includegraphics[scale=0.5]{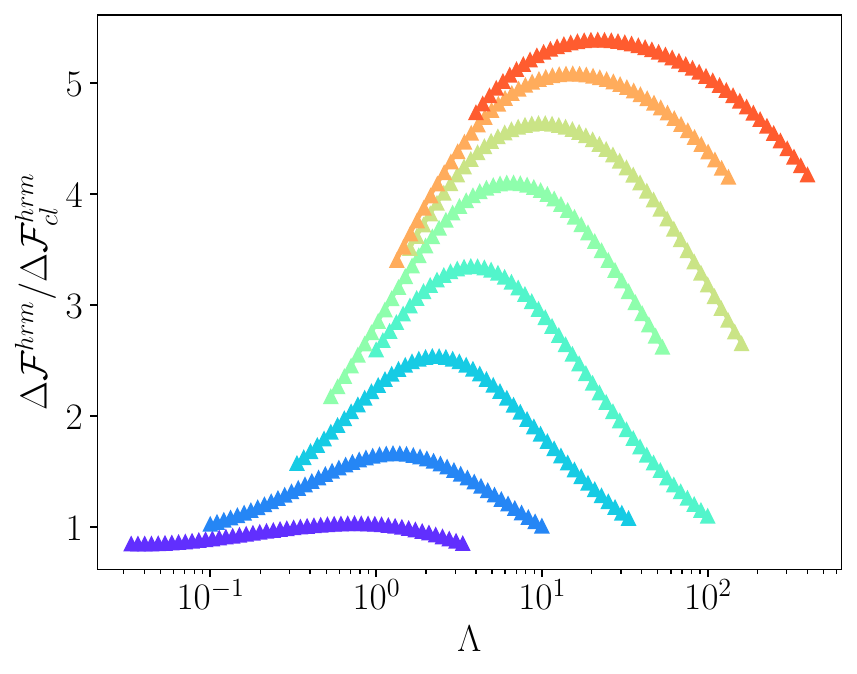}
    \includegraphics[scale=0.5]{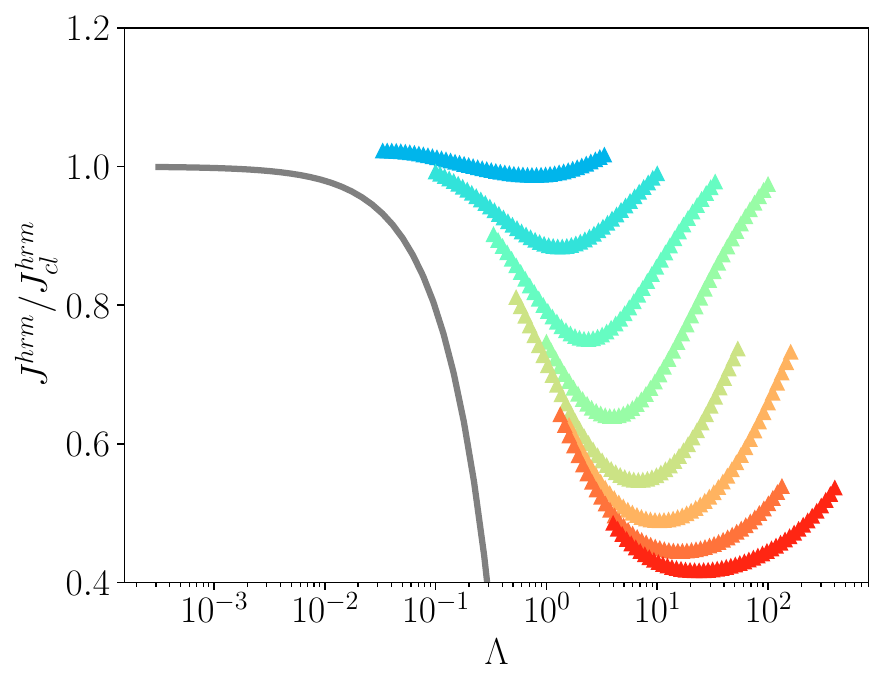}
    \caption{Left: additional data used to identify the maxima of the free energy barrier. Right: Additional data used to identify the maxima of the free energy barrier}
    \label{fig:DF_max_finder}
\end{figure}

\section{Quantitative comparison of the predictions of Eq.~\eqref{eq:rho_eq} against the full numerical solution}

\begin{figure}[h]
\includegraphics[scale=0.5]{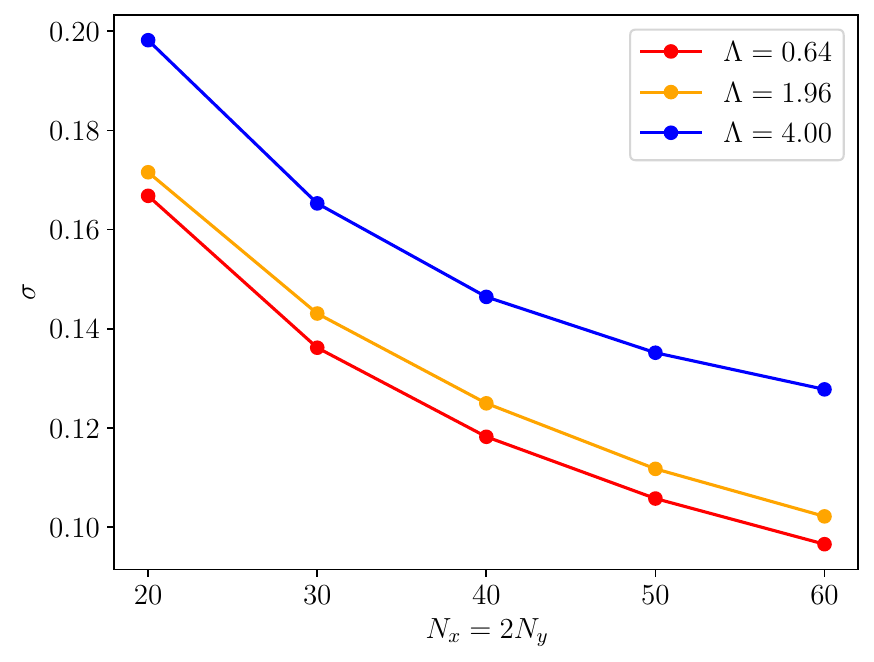}
\caption{Mismatch between the numerical solution and the analytical one (taken as a reference) upon increasing the number of points, $N$, over which the numerical solution is computed.}
\label{fig:err_N}
\end{figure}

\begin{figure}[h]
\includegraphics[scale=0.5]{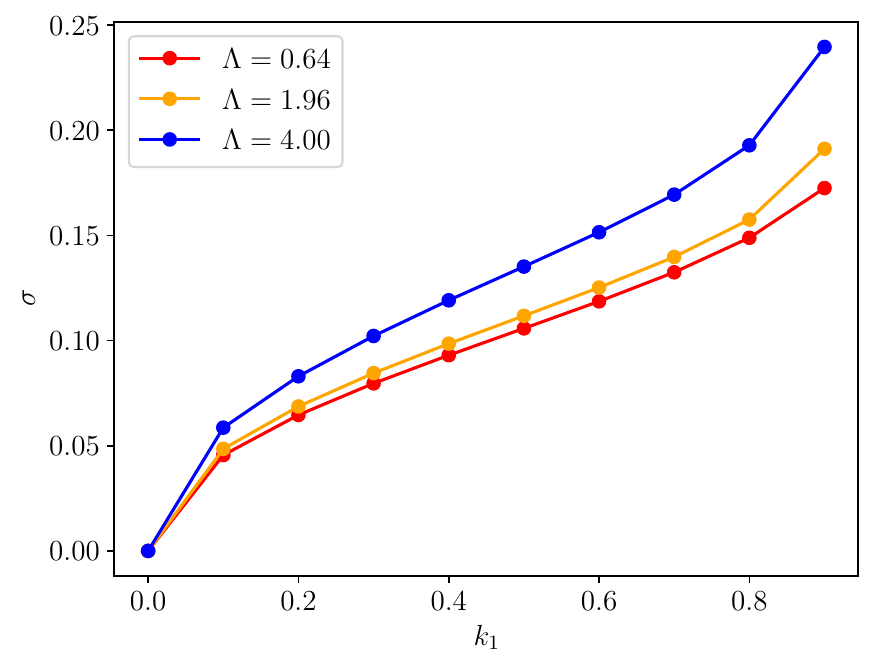}
\includegraphics[scale=0.5]{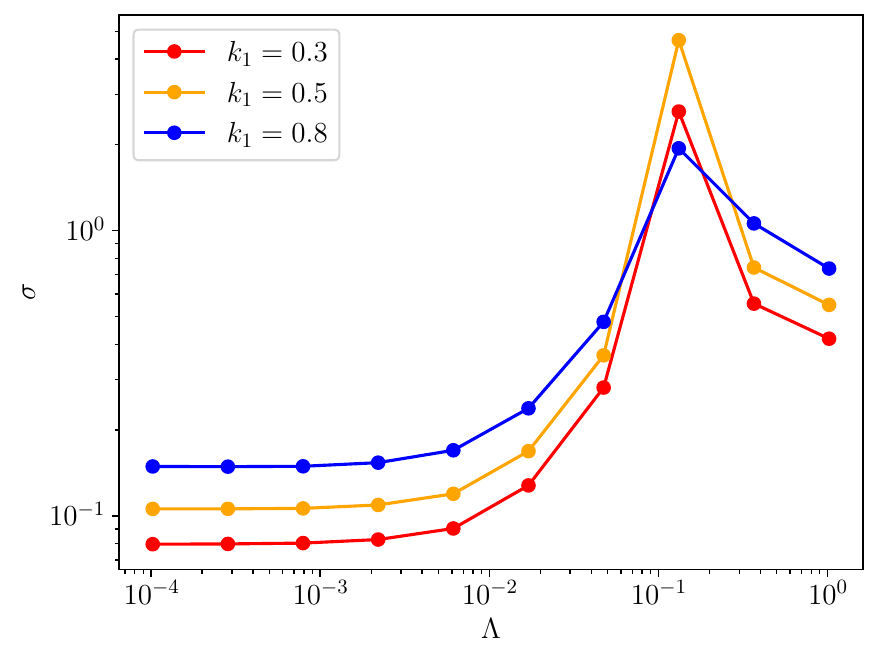}
\caption{Left: mismatch between the numerical solution and the analytical one upon varying $k_1$ for $N=50$ and different values of $\Lambda$ (see legend). Right: mismatch between the numerical solution and the analytical one upon varying $\Lambda$ for $N=50$ and different values of $k_1$ (see legend).}
\label{fig:err_k1}
\end{figure}

In order to assess the reliability of the Fick-Jacobs approach we quantify the mismathc between the two of them as
\begin{align}
\sigma = \int_0^L \dfrac{(\rho_{FJ}(x)-\rho_{num}(x))^2 dx}{\rho_{num}^2(x)}dx\,.
\end{align} 
Since the numerical solutions are computed on a finite grid, firstly we check the convergence of the numerical solution with $N$ by comparing it against the analytical solution, taken as a reference. As shown in Fig.~\ref{fig:err_N}, for $N=\simeq 50$ the dependence of the numerical solution on $N$ becomes very mild and hence in the following we stick to $N=50$.
Next we assess the dependence of $\sigma$ on $k_1$ and $\Lambda$. As shown in Fig.\ref{fig:err_k1}, the analytical prediction is reliable (within $10\%$ error) up to $\Lambda\simeq 0.1$. For larger values of $\Lambda$ the approximations is quantitatively reliable only for smaller values of $k_1$.

\section{Classical Fick-Jacobs}

In a nutshell the Fick-Jacobs approximation deals with non-self-interacting systems whose density, $p$, is governed, in the overdamped regime by the Smoluchowski equation:

\begin{align}
\!\!\!\!\dot{p}(x,y,t)=\nabla\cdot\left[D\nabla p(x,y,t)+Dp(x,y,t)\beta\nabla U(x,y)\right]
\end{align}
where $D$ is the diffusion coefficient and 
\begin{align}
U(x,y)=\begin{cases}
W(x,y) & |y|\leq h(x)\\
\infty & |y|>h(x)
\end{cases}
\end{align}
encodes for a conservative potential, $W$, and for the confinement.
The Fick-Jacobs approximation relies on the following ansatz:

\begin{align}
p(x,y,t)=\rho(x,t)\dfrac{e^{-\beta W(x,y)}}{e^{-\beta A(x)}}
\end{align}
with 
\begin{align}
e^{-\beta A(x)}=\frac{1}{L_y}\int_{-\infty}^{\infty}e^{-\beta W(x,y)}dy
\end{align}
where $L_Y$ is a characteristic length (that is irrelevant for any physically meaning quantity). 
Substituting the ansatz into the Smoluchowski equation leads 
\begin{align}
\dot{\rho}(x,t)=\partial_{x}\left[D\partial_{x}p(x,y,t)+Dp(x,y,t)\beta\partial_{x}A(y)\right]
\end{align}
\end{document}